\documentclass[sigconf,10pt]{acmart}

\renewcommand\footnotetextcopyrightpermission[1]{} 
\makeatletter
\renewcommand\@formatdoi[1]{\ignorespaces}
\makeatother

\usepackage{booktabs} 
\usepackage[inline]{enumitem}
\usepackage[linesnumbered ]{algorithm2e} 
\usepackage{todonotes}
\usepackage{subcaption}
\usepackage[utf8]{inputenc}

\usepackage{breakurl}

\begin{document}
\title[Unveiling the potential of GNN for network modeling and optimization in SDN]{Unveiling the potential of Graph Neural Networks\\for network modeling and optimization in SDN}

\author{Krzysztof Rusek}
\affiliation{%
  \institution{AGH University of Science and Technology, Department of Telecommunications, Krakow, Poland.}
}
\email{krusek@agh.edu.pl}

\author{José Suárez-Varela}
\affiliation{%
  \institution{Universitat Politècnica de Catalunya, Spain}
}
\email{jsuarezv@ac.upc.edu}

\author{Albert Mestres}
\affiliation{%
  \institution{Universitat Politècnica de Catalunya, Spain}
}
\email{amestres@ac.upc.edu}

\author{Pere Barlet-Ros}
\affiliation{%
  \institution{Universitat Politècnica de Catalunya, Spain}
}
\email{pbarlet@ac.upc.edu}

\author{Albert Cabellos-Aparicio}
\affiliation{%
  \institution{Universitat Politècnica de Catalunya, Spain}
}
\email{acabello@ac.upc.edu}

\hyphenation{responsible}

\begin{abstract}

Network modeling is a critical component for building self-driving Software-Defined Networks, particularly to find optimal routing schemes that meet the goals set by administrators. However, existing modeling techniques do not meet the requirements to provide accurate estimations of relevant performance metrics such as delay and jitter. In this paper we propose a novel Graph Neural Network (GNN) model able to understand the complex relationship between topology, routing and input traffic to produce accurate estimates of the per-source/destination pair mean delay and jitter. GNN are tailored to learn and model information structured as graphs and as a result, our model is able to generalize over arbitrary topologies, routing schemes and variable traffic intensity. In the paper we show that our model provides accurate estimates of delay and jitter (worst case $R^2=0.86$) when testing against topologies, routing and traffic not seen during training. In addition, we present the potential of the model for network operation by presenting several use-cases that show its effective use in per-source/destination pair delay/jitter routing optimization and its generalization capabilities by reasoning in topologies and routing schemes not seen during training.
\end{abstract}

\maketitle
\section{Introduction}

\subsection{Motivation}

Network optimization is a well-known and established topic with the fundamental goal of operating networks efficiently. In the context of the SDN paradigm, network optimization is achieved by incorporating two components to the SDN controller: (i) a network model and (ii) an optimization algorithm (e.g, \cite{TEakyildiz}). Typically, the network administrator configures the network policy (goals) in the optimization algorithm that uses the network model to obtain the configuration that meets the goals.  

In this traditional and well-known architecture the model is responsible for predicting the performance (e.g, per-link utilization) of the network (e.,g topology) for a particular configuration (e.g, routing). Then the optimization algorithm is tasked to explore the configurations to find one that meets the goals of the network administrator. An example of this is Traffic Engineering, where the goal is finding a routing configuration that keeps the per-link utilization below the per-link capacity. Since the dimensionality of the configuration is typically very large, efficient optimization strategies reduce them by using expert knowledge. The networking community has developed over decades a large set of network models and optimization strategies~\cite{rexfordOptimization}.

One of the fundamental characteristics of network optimization is that \emph{we can only optimize what we can model}. For example, in order to optimize the jitter of the packets traversing the network we need a model able to understand how jitter relates to other network characteristics. In the field of fixed networks many accurate network models have been developed in the past, particularly using Queuing Theory \cite{queuingModels}. However, such models make some simplifications like assuming some non-realistic properties of real-world networks (e.g., generation of traffic with Poisson distribution, probabilistic routing). Moreover, they do not work well for networking problems involving multi-hop routing (i.e., multi-point to multi-point queueing) and estimation of end-to-end performance metrics \cite{experienceDriven}. As a result, they are not accurate for large networks with realistic routing configurations and as such, delay, jitter and losses remain as critical performance metrics for which no practical model exists.

Recent advances in Artificial Intelligence (AI) \cite{googleNature} have led to a new era of Machine Learning (ML) techniques such as Deep Learning \cite{deepLearning}. This has attracted the interest of the networking community to try to take advantage of these novel techniques to develop a new breed of models, particularly focused on complex network behavior and/or metrics.

In this context, relevant research efforts are being devoted into this new field. Researchers are using neural networks to model computer networks \cite{wangMachineLearning} and using such models for network optimization \cite{IntelligentRouting}, in some cases in combination with advanced strategies based on Deep Reinforcement Learning \cite{learningRouting,experienceDriven,deepRMSA}.

Such proposals \cite{deepQ,mestresModeling} typically use well-known Neural Networks (NN) architectures like fully-connected Neural Networks, Convolutional Neural Networks (extensively used for image processing), Recurrent Neural Networks (used for text processing) or Variational Auto-Encoders. However, computer networks are fundamentally represented as graphs, and such types of NN are \emph{not designed to learn information structured as graphs}. As a result, the models trained are strongly limited: they provide limited accuracy and are unable to generalize in terms of topologies or routing configurations. This is one of the main reasons why ML-based network optimization techniques have -at the time of this writing- failed to meet its expectations and clearly outperform traditional techniques.

\subsection{Objectives}

In this paper we aim to address these issues and we present RouteNet, a pioneering network model based on Graph Neural Networks (GNN)~\cite{graphNetworks}. Our model is able to understand the complex relationship between topology, routing and input traffic to produce accurate estimates of the per-source/des-tination pair mean delay and jitter. GNN are tailored to learn and model information structured as graphs and as a result our model is able to generalize over arbitrary topologies, routing schemes and variable traffic intensity. To the best of our knowledge, this is the first work to address such fundamental networking problem using ML-based techniques able to learn and \emph{generalize}.

Graph Neural Network (GNN) models have grown in popularity in recent years and are particularly designed to operate on graphs with the aim of achieving relational reasoning and combinatorial generalization. In other words, GNNs facilitate the learning of relations between entities in a graph and the rules for composing them (i.e., they have a strong \textit{relational inductive bias} \cite{relationalInductiveBias}). Specifically, our model is inspired by Message-passing Neural Networks \cite{MPNN}, such models are already used in chemistry to develop new compounds. With this framework we design a new model that captures \emph{meaningfully} traffic routing over network topologies. This is achieved by modeling the relationships 
of the links in topologies with the source-destination paths resulting from the routing schemes and the traffic flowing through them.

\subsection{Contributions}

In order to test the accuracy of our model we train it with a dataset generated using a per-packet simulator (Omnet++ \cite{omnet}) resulting in high estimation accuracy of delay and jitter (worst case $R^2=0.86$)  when testing against topologies, routing and traffic not seen during training. More importantly, we verify that our model is able to generalize and for instance, when training the model with samples of a 14-node topology the model is able to provide accurate estimates in a never seen 24-node network.

Finally, and in order to showcase the potential of our model we present a series of use-cases applicable to a SDN architecture:

\begin{enumerate}
\item \textbf{Routing Optimization:} We first show that our model can be used to find routing schemes that minimize per-source/destination average delay and/or jitter. We benchmark it against traditional utilization-aware models (e.g., OSPF) achieving improvements up to 43.5\%. We show that this model can be also used for SLA optimization, where delay or jitter SLA is maintained for a set of source-destination pairs even when the overall traffic volume increases.

\item \textbf{Link failures:} In order to show the generalization capabilities of our model we show that it is able to produce estimates of delay and jitter in the presence of link-failures. That is, changes in the topology and the routing.

\item \textbf{What-if Scenarios:} Finally we show that the model can be used to reason in what-if scenarios answering the following questions: What will happen to the delay/jitter of the network if a new user is added? And, how should I upgrade the network to significantly reduce the overall delay and jitter?
\end{enumerate}

\section{Network architecture}

\begin{figure}[!t]
	\centering
	\includegraphics[width=1.0\linewidth]{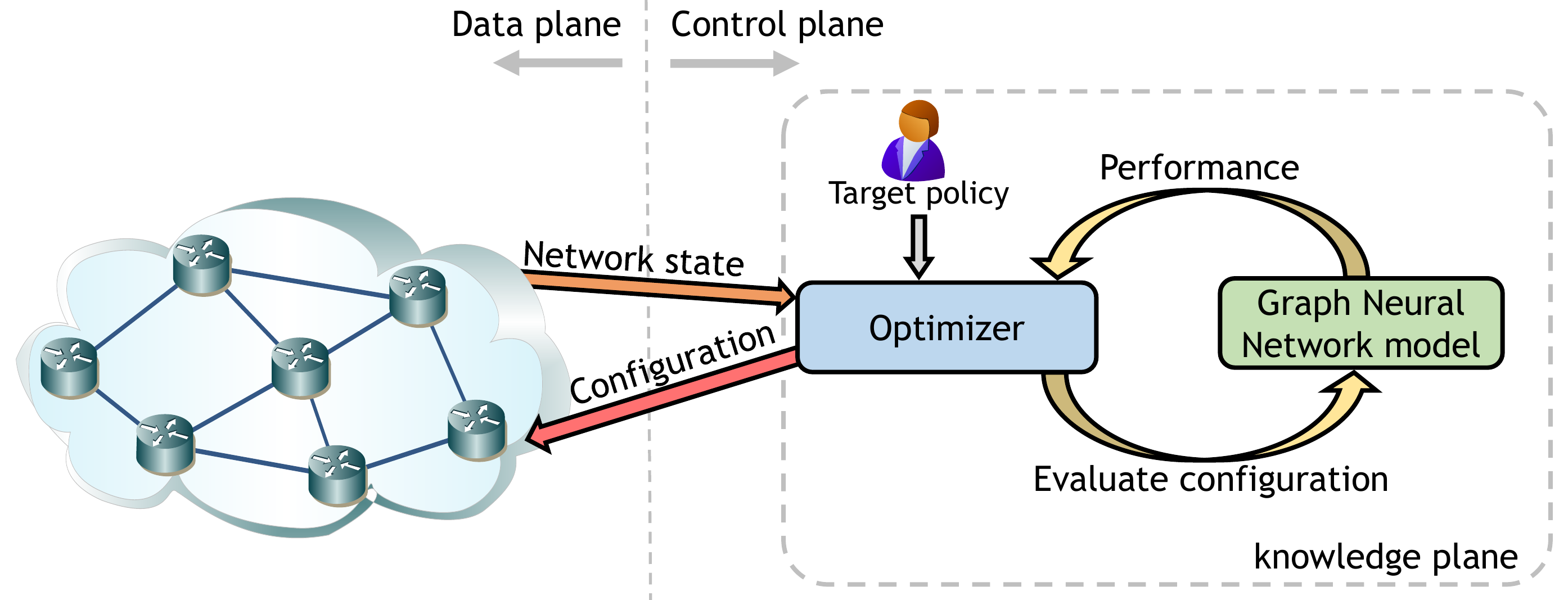}
	\caption{Architecture for network optimization}
	\label{fig:scenario}
	\vspace{-0.2cm}
\end{figure}

Network modeling enables the control plane to further exploit the potential of SDN to perform fine-grained management. This permits to evaluate the resulting performance of what-if scenarios without the necessity to modify the state of the data plane. It may be profitable for a number of network management applications such as optimization, planning or fast failure recovery. For instance, in Fig. \ref{fig:scenario} we show an architecture of a use-case that performs network optimization within the context of the \textit{knowledge-Defined Networking} (KDN) paradigm \cite{kdn}. In this case, we assume that the control plane receives timely updates of the network state (e.g., traffic matrix, delay measurements). This can be achieved by means of ``conventional'' SDN-based measurement techniques (e.g., OpenFlow \cite{openflow}, OpenSketch \cite{opensketch}) or more novel telemetry proposals such as INT for P4 \cite{int-p4} or iOAM \cite{ioam}. Likewise, in the knowledge plane there is an optimizer whose behavior is defined by a given target policy. This policy, in line with intent-based networking, may be defined by a declarative language such as NEMO \cite{nemo} and finally being translated to a (multi-objective) network optimization problem. In this point, an accurate network model can play a crucial role in the optimization process by leveraging it to run optimization algorithms (e.g., hill-climbing) that iteratively explore the performance of candidate solutions in order to find the optimal configuration. We intentionally leave out of the scope of this architecture the training phase.

\begin{figure}[!b]
	\centering
	\vspace{-0.1cm}
	\includegraphics[width=1.0\linewidth]{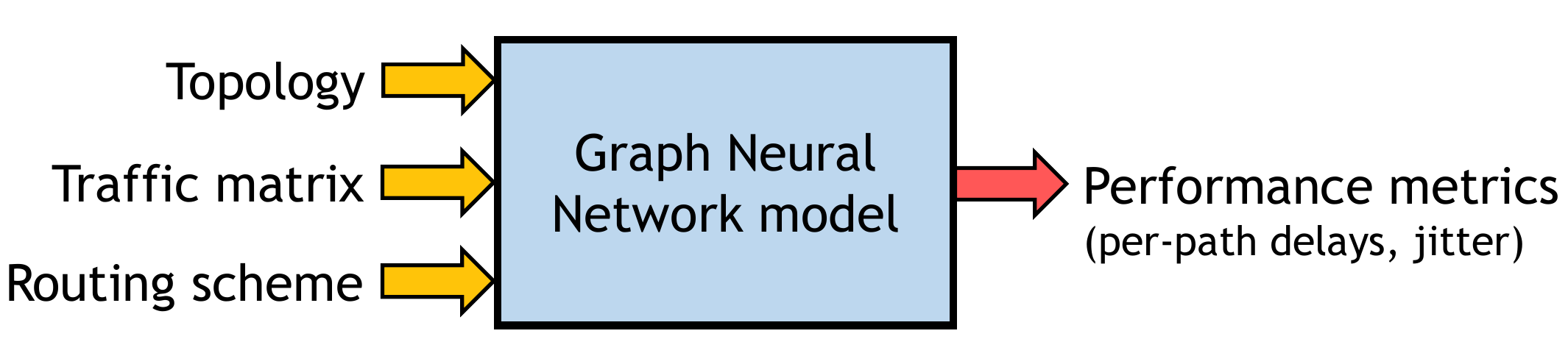}
	\caption{Scheme of RoutNet - our GNN-based model.}
	\vspace{-0.2cm}
	\label{fig:DL-model}
\end{figure}

To be successful in scenarios like the one proposed above, the network model should meet two main requirements: \newline $(i)$ providing accurate results, and $(ii)$ having a low computational cost to allow network optimizers to operate in short time scales. Moreover, it is essential for optimizers to have enough flexibility to simulate what-if scenarios involving different routing schemes, changes in the topology and variations in the traffic matrix. To this end, we rely on the capability of Graph Neural Network (GNN) models to efficiently operate and generalize over environments represented as graphs. Our GNN-based model, RouteNet, inspired by the Message-Passing Neural Network \cite{MPNN} used in the chemistry field, is able to propagate any routing scheme throughout a network topology and abstract meaningful information of the current network state. Fig. \ref{fig:DL-model} shows a schematic representation of the model. More in detail, RouteNet takes as input $(i)$ a given topology, $(ii)$ a source-destination routing scheme (i.e., relations between end-to-end paths and links) and $(iii)$ a traffic matrix (defined as the bandwidth between each pair of nodes in the network), and finally produces performance metrics according to the current network state (e.g., per-path delays or jitter). To achieve it, RouteNet uses fixed-dimension vectors that encode the states of paths and links and propagate the information among them according to the routing scheme.
In Section \ref{sec:use-cases}, we provide some relevant use-cases with experiments that exhibit how we can benefit from this GNN model in different network-related problems.

\section{Network modeling with GNN}
\label{sec:modeling-gnn}

In this section, we provide a detailed mathematical description of RouteNet, the GNN-based model proposed in this paper and designed specifically to operate in networking scenarios.

\subsection{Notation}

A computer network can be represented by a set of links $\mathcal{N}=\{l_i\},\quad i \in (0,1,\ldots,n_l)$,
and the routing scheme in the network by a set of paths $\mathcal{R}=\{p_k\} \quad k \in (0,1,\ldots,n_p) $. Each path is defined as a sequence of links $p_k=(l_{k(0)},\ldots, l_{k(|p_k|)})$, where $k(i)$ is the index of the $i$-th link in the path $k$. The properties (features) of both links and paths are denoted by $\mathbf x_{l_i}$ and $\mathbf x_{p_i}$.

\subsection{Message Passing on Paths}
Let us consider the delay on path $p_k=(l_{k(0)},l_{k(1)},l_{k(2)}\ldots)$.
The state of every link in this path and consequently, the associate delays, depend on all the traffic traversing these links. If packet loss is negligible, the order of links in the path does not matter. Then, the delay could be computed as $\sum_i d(l_{k(i)})$, where $d(l_j)$ represents the delay on the $j$-th link.
However, the presence of links with losses introduces sequential dependence between the link states.

Let the state of a link be described by $\mathbf{h}_{l_i}$, which is an unknown hidden vector.
Similarly, the state of a path is defined by $\mathbf{h}_{p_i}$.
We expect the link state vector to contain some information about the link delay, packet loss rate, link utilization, etc. Likewise, the path state is expected to contain information about end-to-end metrics such as delays or total losses. Considering these assumptions, we can state the following principles:

\begin{enumerate}[label=\arabic*)]
    \item The state of a path depends on the states of all the links in the path.
    \item The state of a link depends on the states of all the paths including the link.
\end{enumerate}

These principles can be matematically formulated with the following expressions:
\begin{gather}
	\mathbf{h}_{l_i} = f(\mathbf{h}_{p_1},\ldots, \mathbf{h}_{p_j}),  \quad l_i \in p_k,\space k=1, \ldots, j \label{eq:hl}\\ 
	\mathbf{h}_{p_k} = g(\mathbf{h}_{l_{k(0)}},\ldots, \mathbf{h}_{l_{k(|p_k|)}}) \label{eq:hp}
\end{gather}
where $f$ and $g$ are some unknown functions. 

It is well-known that neural networks can work as universal function approximators. However, a direct approximation of functions $f$ and $g$ is not possible in this case given that:
$(i)$ Equations~\eqref{eq:hl} and~\eqref{eq:hp} define an implicit function (a nonlinear system of equations with the states being hidden variables),
$(ii)$ these functions depend on the input routing scheme, and
$(iii)$ the dimensionality of each function is very large. This would require a vast set of training samples.

Our goal is to achieve a structure for $f$ and $g$ being invariant for the routing scheme but still being aware of it. For this purpose, we propose RouteNet, a Graph Neural Network architecture based on \emph{message-passing neural networks} (MPNN)~\cite{MPNN}, which were already successfully applied to a quantum chemistry problem.

Algorithm~\ref{alg:mpnn} describes the forward propagation (and the internal architecture) of the network.
In this process, RouteNet receives as input the initial path and link features $\mathbf x_p$, $\mathbf x_l$ and the routing description $\mathcal{R}$, and outputs inferred per-path metrics ($\hat{\mathbf y}_p$). Note that we simplified the notation by dropping sub-indexes of paths and links.

\begin{algorithm}[!t]
	\SetAlgoLined
	\KwIn{$\mathbf x_p$, $\mathbf x_l$,$\mathcal{R}$} \label{lin:in}
	\KwOut{$\mathbf h_p^T$, $\mathbf h_l^T$, $\hat{\mathbf y}_p$}
	\tcp{Initialize states of paths and links}
	\ForEach{$p \in \mathcal{R}$}{$\mathbf h_p^0 \leftarrow [\mathbf x_p,0\ldots, 0]$}\label{lin:hp0}
	\ForEach{$l \in \mathcal{N}$}{$\mathbf h_l^0 \leftarrow [\mathbf x_l,0\ldots, 0]$}\label{lin:hl0}
	\For{$t=1$ to $T$}{\label{lin:forT}
	    \tcp{Message passing from links to paths}
		\ForEach{$p \in \mathcal{R}$}{
			\ForEach{$l \in p$}{\label{lin:mp}
				$\mathbf h_p^{t} \leftarrow RNN_t(\mathbf h_p^t,\mathbf h_l^t)$ \label{lin:hp} \\
				$\tilde{\mathbf m}_{p,l}^{t+1}\leftarrow \mathbf h_p^{t} $ \label{lin:up}
			}
			$\mathbf h_p^{t+1} \leftarrow \mathbf h_p^{t}$
		}
		\tcp{Message passing from paths to links}
		\ForEach{$l \in \mathcal{N}$}{
			$\mathbf m_l^{t+1} \leftarrow \sum_{p:l \in p} \tilde{\mathbf m}_{p,l}^{t+1} $ \label{lin:ml}\\
			$ \mathbf h_l^{t+1} \leftarrow U_t\left(\mathbf h_l^t,\mathbf m_l^{t+1}\right)$ \label{lin:ul}
		}
	}
	\tcp{Readout function}
	$\hat{\mathbf y}_p \leftarrow F_p(\mathbf h_p)$\label{lin:read}
	\caption{Internal architecture of RouteNet}
	\label{alg:mpnn}
\end{algorithm}

RouteNet's architecture enables dealing with the circular dependencies described in equations \eqref{eq:hl} and \eqref{eq:hp}, and supporting arbitrary routing schemes (which are inherently represented within the architecture). This is all thanks to the ability of GNNs to address problems represented as graphs and solve circular dependencies by making an iterative approximation to fixed point solutions.

In order to address the circular dependencies, RouteNet repeats the same operations over the state vectors $T$ times (loop from line~\ref{lin:forT}). These steps represent the convergence process to the fixed point of a function from the initial states $\mathbf{h}_{p}^{0}$ and $\mathbf{h}_{l}^{0}$.

Regarding the issue of routing invariance (more generically known as topology invariance in the context of graph-related problems). This requires the use of a structure that is able to represent graphs of different topologies and sizes. In our case, we aim at representing different routing schemes in a uniform way. One state-of-the-art solution for this problem~\cite{rusek2018message} proposes using neural message passing architectures that combine both: a representation of the topology as a graph, and vectors to encode the link states. In this context, RouteNet can be interpreted as an extension of a vanilla message passing neural network that is specifically suited to represent the dependencies among links and paths given a routing scheme (Equations \eqref{eq:hl} and \eqref{eq:hp}).

In Algorithm~\ref{alg:mpnn}, the loop from line~\ref{lin:mp} and the line~\ref{lin:ml} represent the \emph{message-passing} operations that exchange mutually the information encoded (hidden states)  among links and paths. Likewise, lines \ref{lin:up} and~\ref{lin:ul} are \emph{update} functions that encode the new collected information into the hidden states.
The path update (line \ref{lin:up}) is a simple assignment, while the link update (line \ref{lin:ul}) is a trainable neural network. In general, the path update could be also a trainable neural network.

From a computational point of view, the loops over links and paths are the most expensive parts of the algorithm. An upper bound estimate of complexity is $O(n^3)$, where $n$ is the number of nodes in the network. 
This assumes the worst case scenario, where all the paths have length $n$. 
Typically, the expected diameter in real networks is around $\log(n)$ (e.g., Erd˝os–R´enyi random graphs), thereby the complexity would decrease to $O(n^2\log(n))$.

This architecture provides flexibility to represent any source-destination routing scheme. This is achieved by the direct mapping of $\mathcal{R}$ (i.e., the set of end-to-end paths) to specific message passing operations among link and path entities that define the architecture of RouteNet. Thus, each path collects messages from all the links included in it (loop from line~\ref{lin:mp}) and, similarly, each link receives messages from all the paths containing it (line~\ref{lin:ml}). Given that the order of paths traversing the same link does not matter, we used a simple summation for the path-level message aggregation. However, in the case of links, the presence of packet losses may imply sequential dependence in the links that form every path. Consequently, we use a Recurrent Neural Network (RNN) for the link-level message aggregation. Note that RNNs are well suited to capture dependence in sequences of variable size (e.g., text processing). This allows us to model losses in links and propagate this information through all the paths. For an input sequence $\mathbf i_1,\mathbf i_2, \ldots$ and an initial hidden state $\mathbf s_0$, the output of a RNN is defined as:

\begin{equation*}
    (\mathbf o_{t}, \mathbf s_{t}) = RNN(\mathbf s_{t-1},\mathbf i_{t}). 
\end{equation*}

In our case, we use a simple version of a RNN, where $o_t=s_t$.

Moreover, the use of these message aggregation functions (RNN and summation) enables to significantly limit the dimensionality of the problem. The purpose of these functions is to collect an arbitrary number of messages received in every (link or path) entity, and compress this information into fixed-dimension arrays (i.e., hidden states). Note that the size of the hidden states of links and paths are configurable parameters. In the end, all the hidden states in RouteNet represent an explicit function containing information of the link and path states. This enables to leverage them to infer various features at the same time. Given a set of hidden states $\mathbf h_p^T$ and  $\mathbf h_l^T$, it is possible connect readout neural networks to estimate some path and/or link-level metrics. This can be typically achieved by using ordinary fully-connected networks with some layers and proper activation functions. In Algorithm~\ref{alg:mpnn}, the function $F_p$ (line~\ref{lin:read}) represents a readout function that predicts some path-level features ($\hat{\mathbf y}_p$) using as input the path hidden states $\mathbf h_p$. Similarly, it would be possible to infer some link-level features ($\hat{\mathbf y}_l$) using information from the link hidden states $\mathbf h_l$.

\subsection{Delay model}
\label{subsec:delay-model}
RouteNet is a general neural architecture capable of modeling various network performance metrics. In order to apply it to particular problems, the following design decisions may be considered:
\begin{enumerate*}[label=\textbf{\arabic*})]
\item The size of the hidden states for both paths ($\mathbf h_p$) and links ($\mathbf h_l$).
\item The number of message passing iterations ($T$).
\item The neural network architectures for $RNN$, $U$, and $Fp$.
\end{enumerate*}
The last decision may be the most complex one, since there are multiple types of neural networks and possible configurations. In our particular case, where we use RouteNet to model per-path delays, we decided to use Gated Recurrent Units (GRU) for both $U$ and $RNN$. The reason behind this, is that GRUs are simpler than LSTM networks (i.e., there is no output gate) and \textit{a priori} can achieve comparable performance~\cite{Chung14a}. GRUs are recurrent units that have an internal structure that by design reuses weights (i.e., weight tying), which considerably simplifies the model.

We modeled the readout function ($Fp$) with a fully-con-nected neural network and use SELU activation functions in order to achieve desirable scaling properties~\cite{Klambauer2017}. These hidden layers are interleaved with two dropout layers.

The dropout layers play two important roles in the model. During training, they help to avoid overfitting, and during the inference, they can be used for Bayesian posterior approximation~\cite{Gal2015}. This allows us to asses the confidence of the network predictions and avoid the issue of adversarial examples~\cite{Goodfellow2014ExplainingAH}. Typically, when a neural network is optimized to minimize the error for a particular output, the solution may be too optimistic. Thus, repeating an inference multiple times with random dropout can provide a probabilistic distribution of results, and this distribution (e.g., the spread) can be used to estimate the confidence of the predictions.

\subsection{Jitter model}
\label{subsec:jitter-model}

In order to assess the ability of RouteNet to generalize to different network metrics, we took a model in an early training stage that produces delay estimates and trained it to produce per-path jitter estimates. The main difference between these two metrics is in the scaling factor, since they are closely related but the jitter spans on different range than the delay.

\section{Evaluation of the accuracy of the GNN model}
In this section, we evaluate the accuracy of RouteNet (Sec. \ref{sec:modeling-gnn}) to estimate the per-source/destination mean delays and jitter in different network topologies and routing schemes.

\subsection{Simulation setup}
\label{sec:simulation-setup}

In order to build a ground truth to train and evaluate the GNN model, we implemented a custom-built packet-level simulator with queues using OMNeT++ (version 4.6)~\cite{omnet}. In this simulator, the delay and jitter modeled in each queue are related to the bandwidth capacity of the corresponding egress links. For each simulation, we measure the average end-to-end delay and jitter experienced during 16k units of time by every pair of nodes. We model the traffic matrix ($\mathcal{TM}$) for each S-D pair in the network as:

\centerline{$\mathcal{TM}(S_i,D_j)~=~\mathcal{U}(0.1, 1)*TI/(N-1)\quad \forall~i,j \in nodes, i\neq j$}

Where $\mathcal{U}(0.1, 1)$ represents a uniform distribution in the range [0.1, 1], \textit{TI} represents a parameter to tune the overall traffic intensity in the network scenario and \textit{N} is the number of nodes in the network.

To train and evaluate the model, we used the 14-node and 21-link NSF network topology~\cite{nsfnet}. Moreover, we use the 24-node Geant2 topology~\cite{geant2} and the 17-node German Backbone Network (GBN)~\cite{gbn} only for evaluation purposes. For the sake of simplicity, we consider the same capacity for all the links in these networks and vary the traffic intensity in each scenario.

\vspace{-0.2cm}
\subsection{Training and Evaluation}
We implemented the RouteNet models of delay and jitter in TensorFlow. The source code and all the training/evaluation datasets used in this paper are publicly available at~\cite{kdngit}.

We trained both models (delay and jitter) on a collection with 260,000 training samples from the NSF network generated with our packet-level simulator. Despite this dataset only contains samples from single topology, it includes around 100 different routing schemes and a wide variety of traffic matrices with different traffic intensity. For the evaluation, we use 30,000 samples.

In our experiments, we select a size of 32 for the path's hidden states ($\mathbf h_p$) and 16 for the link's hidden states ($\mathbf h_l$). The initial path features ($\mathbf x_p$) are defined by the bandwidth that each source-destination path carries (extracted from the traffic matrix). In this case, we do not add initial link features ($\mathbf x_l$). Note that, for larger networks, it might be necessary to use larger sizes for the hidden states. Moreover, every forward propagation we execute $T$=8 iterations. The Dropout rate is equal to 0.5, this means that each training step we randomly deactivate half of neurons in the readout neural network. This also allows us to make a probabilistic sampling of results and infer the confidence of the estimates.

During the training we minimized the mean squared error ($\mathit{MSE}$) between the prediction of RouteNet and the ground truth plus the $L2$ regularization loss ($\lambda =0.1$). The loss function is minimized using an Adam optimizer with an initial learning rate of 0.001. This rate is decreased to 0.0003 after 60,000 training steps approximately.

\begin{figure}[!t]
	\centering
	\input{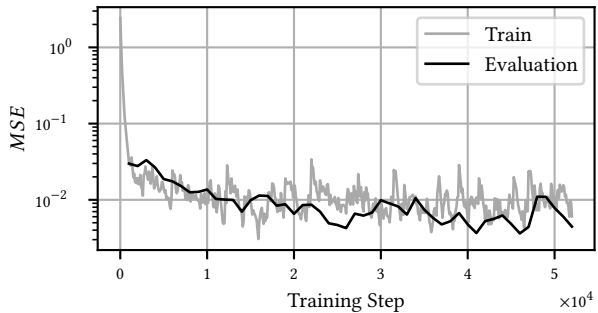}
	\caption{Exponentially smoothed $\mathit{MSE}$ for the delay RouteNet model}
	\vspace{-0.2cm}
	\label{fig:loss}
\end{figure}

We executed the training over 300,000 batches of 32 samples randomly selected from the training set. In our testbed with a GPU Nvidia Tesla K40 XL, this took around 96 hours ($\approx$ 27 samples per second). Figure~\ref{fig:loss} shows the loss during the training process. Here, we observe that the training is stable and the loss drops quickly.

Table~\ref{tab:eval} shows a summary of all the experiments we made in the three different network topologies. We report two statistics: $(i)$ the Pearson correlation $\rho$ and $(ii)$ the percentage of variance explained by the model ($R^2$). For the Geant2 and GBN topologies, we tested the accuracy over a dataset with 100,000 samples. For the NSF network, we utilize the same 30,000 samples used for evaluation during the training process. To calculate the statistics in Table~\ref{tab:eval}, we compute for each sample in the evaluation dataset 50 independent predictions using random dropout and take the median value.
Note that \emph{the Geant2 and GBN networks were never included in the training.} The model was only trained with samples from the NSF network (14 nodes). The high accuracy on Geant2 (24 nodes) and GBN (17 nodes) networks reveals the ability of RouteNet to well generalize even to larger networks.

\begin{table}[!t]
	\caption{Summary of the obtained evaluation results}

	\label{tab:eval}
	\centering
	\begin{tabular}{l|ll|ll|ll}
		\toprule
		& \multicolumn{2}{l|}{\bfseries NSF} & \multicolumn{2}{l|}{\bfseries Geant2}    & \multicolumn{2}{l}{\bfseries GBN}       \\ 
		\hline
		& \bfseries Delay      & \bfseries Jitter      & \bfseries Delay & \bfseries Jitter & \bfseries Delay & \bfseries Jitter \\ 
		\midrule
		$R^2$  & 0.99           & 0.98            &  0.97     &  0.86  & 0.99 & 0.97\\
		$\rho$ & 0.998           & 0.993          &  0.991     & 0.942 & 0.997 & 0.987 \\      
		\bottomrule
	\end{tabular}
\end{table}

\begin{figure}[!t]
	\begin{subfigure}[c]{\linewidth}
    	\centering
    	\includegraphics[width=0.97\linewidth]{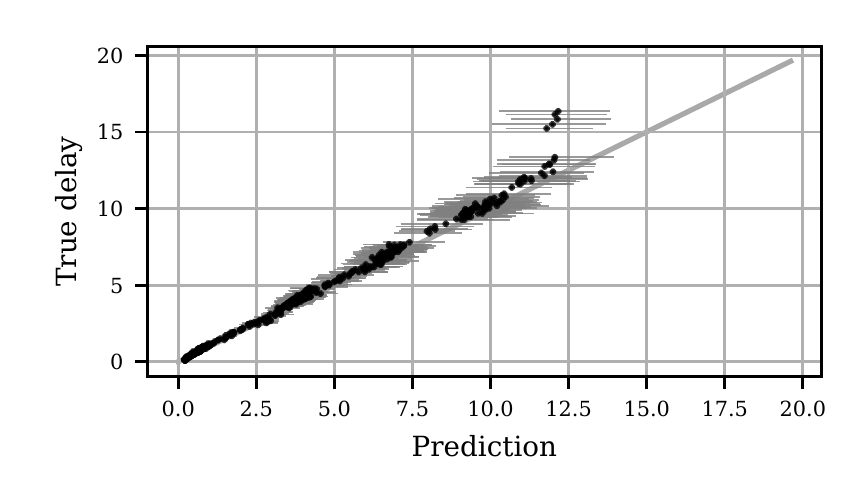}
    	\vspace{-0.4cm}
    	\caption{Geant2}
    	\label{fig:regplot_geant}
	\end{subfigure}
    \begin{subfigure}[c]{\linewidth}
        \centering
    	\includegraphics[width=0.97\linewidth]{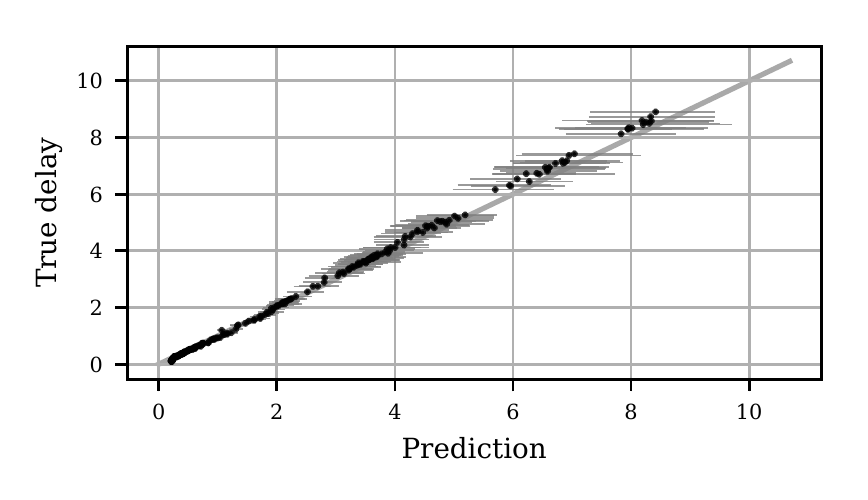}
    	\vspace{-0.4cm}
    	\caption{GBN network}
    	\label{fig:regplotgbn}
    \end{subfigure}
    \vspace{-0.2cm}
	\caption{Regression plots with uncertainty.}
	\label{fig:regplot}
\end{figure}

This generalization capability is partly thanks to the Bayesi-an nature of the network (i.e., the use of layers with random dropout). Figure~\ref{fig:regplot} shows an example of the probabilistic prediction for a single sample of the dataset of Geant2 (Fig. \ref{fig:regplot_geant}) and GBN (Fig. \ref{fig:regplotgbn}).
The dots represent the median value of the predictions of RouteNet, while gray lines show the range containing 95\% of the results.

\begin{figure}
	\centering
	\input{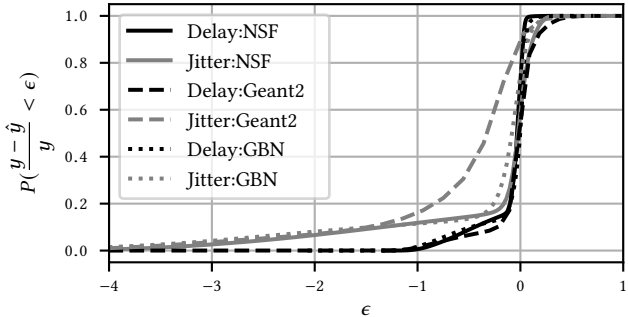}
	\vspace{-1cm}
	\caption{Cumulative Distribution Function of relative error. Solid line for NSF, dashed line for Geant2, dotted line for GBN. $y$ is the true value while $\hat y$ denotes the model prediction.}
	\label{fig:eval_all}
\end{figure}

Statistics like $\rho$ or $R^2$ provide a good picture of the general accuracy of the model. However, there are more elaborated methods that offer a more detailed description of the model behavior. One alternative to gain insight into prediction models is to provide regression plots with the evaluation results. Thus, in Figure~\ref{fig:regplot} we present relevant regression plots for specific scenarios of the evaluation in Geant2 (Fig.~\ref{fig:regplot_geant}) and GBN (Fig.~\ref{fig:regplotgbn}). However, it is not functional showing a regression plot with all the points predicted by RouteNet in all the evaluation scenarios (dozens of millions of points). Hence, we focus on the distribution of residuals (i.e., the error of the model). Particularly, we present a CDF of the relative error (Fig.~\ref{fig:eval_all}) over all the evaluation samples. This allows us to provide a comprehensive view of the whole evaluation in a single plot. In these results, we can observe that the prediction error in general is considerably low. Moreover, we see that the jitter model is more biased compared to the delay model. This can be explained by the fact that this model was trained from a model previously trained for the delay (Sec.~\ref{subsec:jitter-model}), while the delay model was optimized from the beginning to predict delays. Nevertheless, this shows the feasibility to adapt pre-trained models optimized for a specific metric to predict different network performance metrics (i.e., transfer learning~\cite{tranferLearning}).

\subsection{Generalization Capabilities}

This section discusses the generalization capabilities and limitations of RouteNet. As in all ML-based solutions, RouteNet is expected to provide more accurate inference as the distribution of the input data is closer to the distribution of training samples. In our case, it involves topologies with similar number of nodes and distribution of connectivity, routing schemes with similar patterns (e.g., variations of shortest path) and similar ranges of traffic intensities. We experimentally observe the capability of RouteNet to generalize from a 14-node network up to a 24-node network while still providing accurate estimates. Finally, and in order to expand the generalization capabilities of RouteNet, an extended training set must be used including a wider range of distributions of the input elements.

RouteNet's architecture is built to estimate path-level metrics using information from the output path-level hidden states. However, it is relatively easy to change the architecture and use information encoded in the link-level hidden states to produce link-related metrics inference (e.g., congestion probability on links). In addition, the implementation of RouteNet at the time of this writing does not support topologies with different link capacity. Similarly, we made experiments involving different link utilization, which implies comparable complexity. In order to support various per-link capacity, this should be encoded in the initial hidden states of links.

\section{Use-cases}
\label{sec:use-cases}

In this section, we present some use-cases to illustrate the potential of RouteNet (Section \ref{sec:modeling-gnn}) to be used in relevant network optimization tasks. In these use-cases, we use the delay (Section \ref{subsec:delay-model}) and jitter (Section \ref{subsec:jitter-model}) models of RouteNet to evaluate the resulting performance after applying different network configurations. Particularly, we limit the optimization problem to evaluate a given set of candidate configurations (e.g., routing schemes) and select the one that results in better performance according to a given target policy. We compare the performance achieved by our optimizer using the delay/jitter estimates of RouteNet to the results obtained by the same optimizer using measurements of the links' utilization. As a reference, we also provide the results obtained when applying a traditional Shortest Path routing policy. Note that more elaborated state-of-the-art optimization strategies (e.g., \cite{defo}) could (and most likely will) result in better performance than these baselines and also could be combined with RouteNet to further improve the resulting performance. However, we leave the analysis of such techniques out of the scope of this paper, since the purpose of this section is to show the added value of using the lightweight and accurate delay/jitter models provided by RouteNet to perform delay/jitter-aware network optimization.

\begin{figure}[!t]
    \centering
    \begin{subfigure}[b]{1.0\linewidth}
	\includegraphics[width=1.0\linewidth]{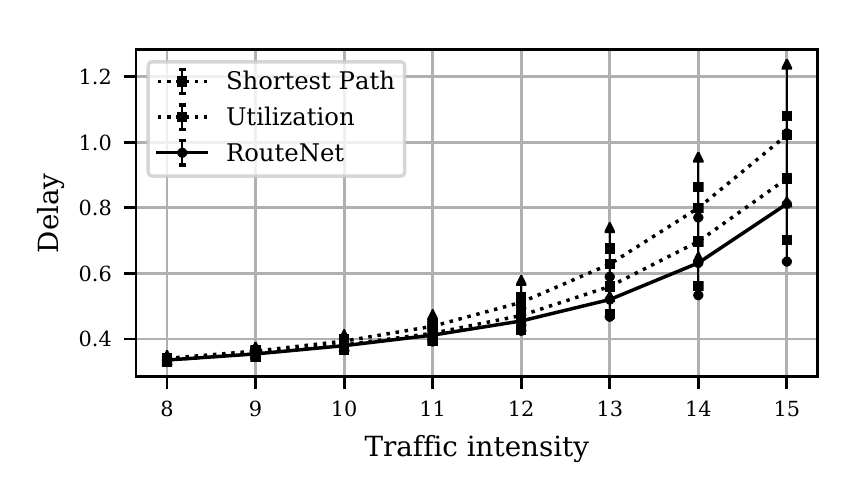}
	    \vspace{-0.65cm}
        \caption{Optimization of mean delay}
	\label{fig:mean_delay}
    \end{subfigure}
    \begin{subfigure}[b]{1.0\linewidth}
	\includegraphics[width=1.0\linewidth]{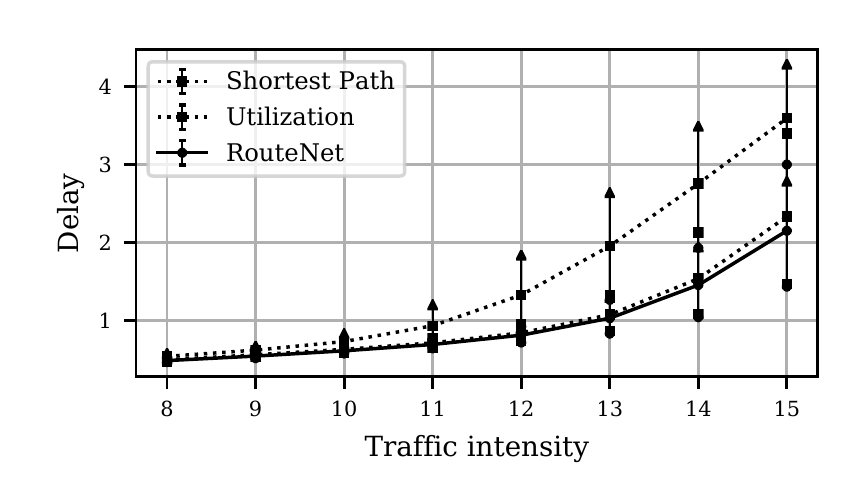}
	    \vspace{-0.65cm}
        \caption{Optimization of maximum delay}
	\label{fig:max_delay}
    \end{subfigure}
    \vspace{-0.5cm}
    \caption{Delay-based optimization}\label{fig:delay}
    \vspace{-0.2cm}
\end{figure}

In this context, using state-of-the-art delay/jitter models to perform online network optimization is typically unfeasible since these models often result into inaccurate estimation (e.g., theoretical models) and/or prohibitive processing cost (e.g., packet-level simulators). All the evaluation in this section is carried in the NSF network topology.

\subsection{Delay/jitter-based routing optimization}
\label{sec:use-cases:1}

In this use-case, the objective is to optimize multiple Key Performance Indicators (KPI) of the network. In particular, we made different experiments where the optimizer aims to: $(i)$ minimize the mean end-to-end delay and jitter, and $(ii)$ minimize the maximum delay and jitter experienced among all the source-destination pairs.

We compare the optimal routing policy obtained by Route-Net with two traditional approaches: $(i)$ Shortest Path (SP) routing, where we compute different SP schemes using the Dijkstra algorithm, and $(ii)$ a more elaborated routing optimizer whose objective is to minimize the bandwidth utilization. This latter strategy represents an upper-bound of the results that could be obtained by traditional routing optimizers based on links' utilization. In particular, for the case of minimizing the mean delay/jitter, we select the routing scheme with minimum mean utilization. In the case of minimizing the maximum delay/jitter, we select the scheme that minimizes the utilization of the link more loaded.

We evaluated the performance obtained by all the routing approaches varying the traffic intensity. Moreover, for a fair comparison, all of them perform the optimization over the same set with 100 different routing schemes randomly generated.

Fig. \ref{fig:mean_delay} shows the minimum mean delay obtained w.r.t. the traffic intensity. Note that traffic intensities (in the x-axis) are in TI units according to the expression in Section \ref{sec:simulation-setup} to generate traffic matrices ($\mathcal{T}$). Moreover, for each traffic intensity, we randomly generated 100 different traffic matrices (TMs) with various per-source/destination traffic distributions. The lines show the average results over the experiments (with those 100 TMs) and the error bars represent the 20/80 percentiles. Likewise, in Fig. \ref{fig:max_delay} we show the results for the use-case where all routing techniques aim at minimizing the maximum end-to-end delay.

\begin{figure}[!t]
    \centering
    \begin{subfigure}[b]{1.0\linewidth}
	\includegraphics[width=1.0\linewidth]{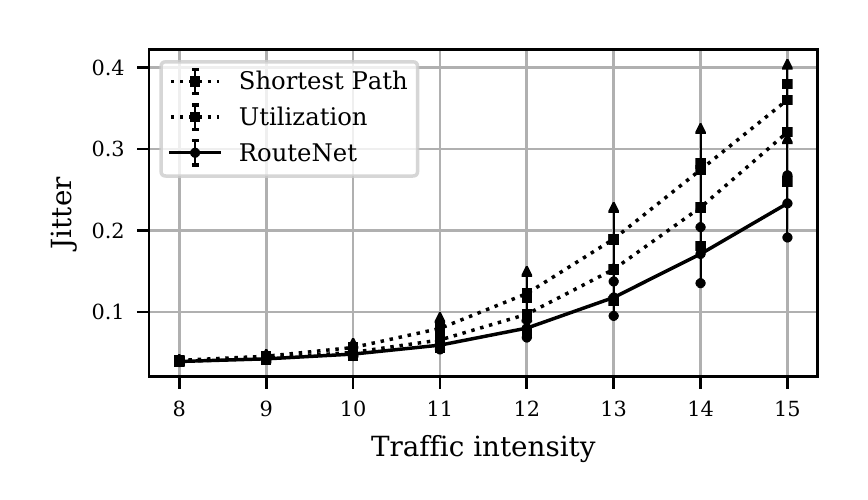}
        \caption{Optimization of mean jitter}
	\label{fig:mean_jitter}
    \end{subfigure}
    \begin{subfigure}[b]{1.0\linewidth}
	\includegraphics[width=1.0\linewidth]{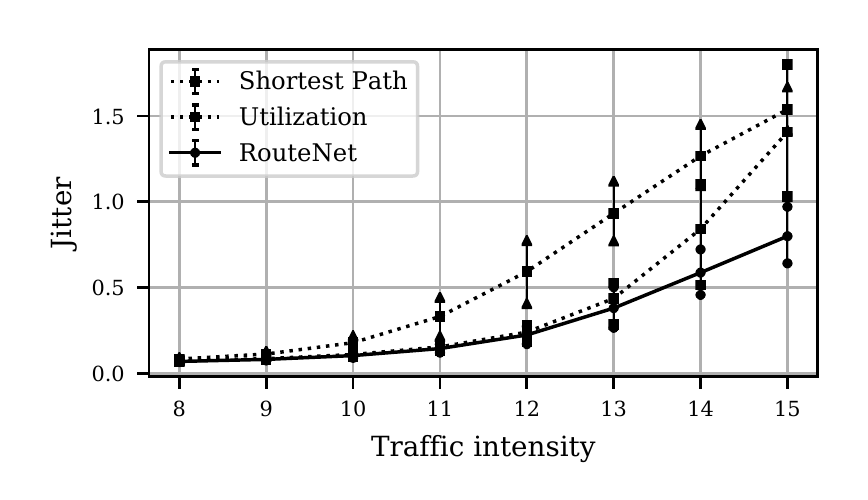}
        \caption{Optimization of maximum jitter}
	\label{fig:max_jitter}
    \end{subfigure}
    \caption{Jitter-based optimization}\label{fig:jitter}
    \vspace{-0.5cm}
\end{figure}

The same experiments were made to evaluate the results optimizing the mean (Fig. \ref{fig:mean_jitter}) and the maximum (Fig. \ref{fig:max_jitter}) jitter experienced by the source-destination pairs in the network.

Considering these results, we can see that, as expected, the performance achieved by the different routing techniques does not differ with low traffic intensity (TI<9). However, the optimizer based on RouteNet delay estimations begins to achieve better performance with medium traffic intensity (TI=10-13) and, for high traffic intensity (TI=13-15), it achieves considerable higher performance. Particularly, with TI=15, it obtains the following results:
\begin{itemize}
    \item When optimizing the mean delay/jitter, the RouteNet-based optimizer achieves 20.87\%/35.27\% lower delay/jit-ter than the SP policy, and 12.18\%/27.21\% lower delay/jitter than the utilization-based optimizer. 
    
    \item When optimizing the maximum delay/jitter, the Route-Net-based optimizer achieves 40.08\%/48.09\% lower delay/jitter than the SP policy, and 8.11\%/43.53\% lower delay/jitter than the utilization-based optimizer.
\end{itemize}

\subsection{SLA optimization}

This use-case represents a network scenario where the routing optimizer must comply a Service Level Agreement (SLA) for some specific clients, while minimizing the impact on the performance of the rest of users in the network. In particular, we consider 4 source-destination pairs to have specific delay requirements. We made the experiments in the NSF network and selected the following source-destination pairs (S-D pairs) that must comply a certain delay requirement: (0,3) (3,4) (3,5) (3,6). Then, the objective is to optimize the routing configuration to guarantee that the traffic among those sources and destinations is below the target delay.

\begin{figure}[!t]
    \centering
    \begin{subfigure}[b]{1.0\linewidth}
	\includegraphics[width=1.0\linewidth]{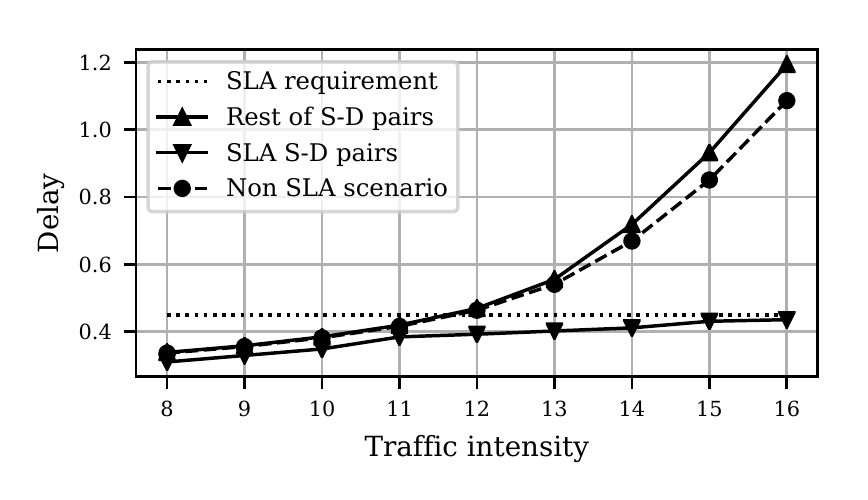}
        \caption{Optimization of mean delay}
        \label{fig:mean_SLA}
    \end{subfigure}
    \begin{subfigure}[b]{1.0\linewidth}
	\includegraphics[width=1.0\linewidth]{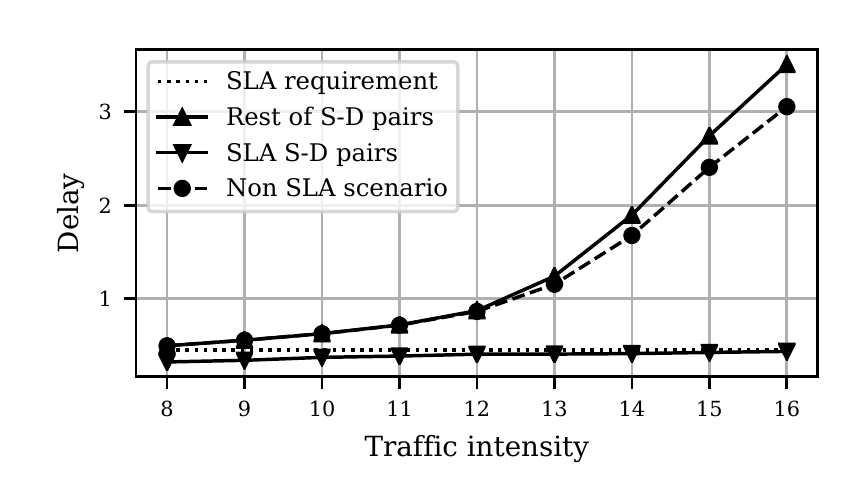}
        \caption{Optimization of maximum delay}
        \label{fig:max_SLA}
    \end{subfigure}
    \caption{Delay optimization under SLA guarantees}\label{fig:sla}
\end{figure}

Fig. \ref{fig:mean_SLA} shows the results in the case that the RouteNet-based optimizer aims to optimize the mean delay experienced in the network, while Fig. \ref{fig:max_SLA} shows the results for the case of minimizing the maximum delay for all the source-destination pairs. In these figures, the dashed line (labeled as ``Non SLA scenario'') represents the results if the optimizer does not distinguish between different traffic classes, and the solid lines represent the results after applying the optimal routing scheme that complies the SLA of the 4 S-D pairs. The dotted line represents the delay requirement of the S-D pairs with SLA, which is an input parameter of the optimizer. Then, we can observe that for the optimization case that considers the SLAs, the delay experienced by the 4 S-D pairs with SLA (labeled as ``SLA S-D pairs'') fulfills the delay requirements (dotted line) even with high traffic intensities (TI=13-16). Moreover, we observe that the rest of S-D pairs without SLA requirements (labeled as ``Rest of S-D pairs'') do not experience a great increase in the mean/maximum delays compared to the ``non SLA scenario''. For instance, with high traffic intensity (TI=15), in the case of optimizing the mean delay, the rest of the traffic only experiences an increase of 9.9\% in the average delay (14.8\% in the case of optimizing maximum delay).

\subsection{Robustness against links failures}

In this use-case, we show how our model is able to generalize in the presence of link failures. When a certain link fails, it is necessary to find a new routing that avoids this link to reroute the traffic. As the number of links failures increases, less paths are available and the network becomes more saturated.

\begin{figure}[!t]
    \centering
    \begin{subfigure}[b]{1.0\linewidth}
	\includegraphics[width=1.0\linewidth]{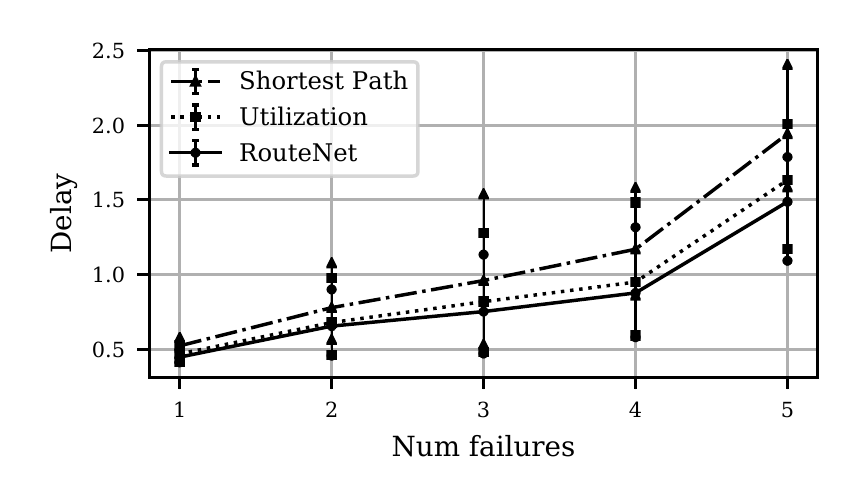}
        \caption{Mean delay}
	\label{fig:mean_failures}
    \end{subfigure}
    \begin{subfigure}[b]{1.0\linewidth}
	\includegraphics[width=1.0\linewidth]{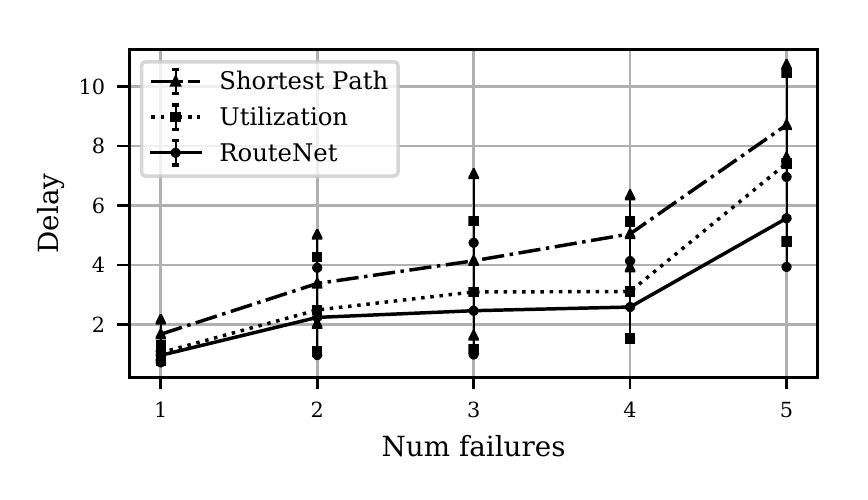}
        \caption{Maximum delay}
	\label{fig:max_failures}
    \end{subfigure}
    \caption{Delay optimization under the presence of different link failures}\label{fig:failures}
    \vspace{-0.4cm}
\end{figure}

We evaluate the performance of the aforementioned methods under the presence of link failures following the same methodology than in the first use-case (see Sect.~\ref{sec:use-cases:1}). The initial network state is a low traffic intensity scenario (TI=8).

Fig.~\ref{fig:failures} shows the optimized mean delay (Fig.~\ref{fig:mean_failures}) and the optimized max delay (Fig.~\ref{fig:max_failures}).
Each point in the graph corresponds to the optimal delay obtained under 10 random possible links failures. We observe that, as shown in the first use-case, the mean and the maximum delays increase as the network is more congested and, in these scenarios, the RouteNet-based optimization mechanism outperforms traditional approaches.

\subsection{What-if scenarios}

One application of interest of network modeling is that network operators can simulate hypothetical what-if scenarios to evaluate the resulting performance before making strategic decisions. These decisions, for instance, include making agreements to route a considerable bulk of traffic from other network (e.g., BGP peering agreements) or finding a network upgrade that results more beneficial given a limited budget.

\begin{figure}[!t]
    \centering
    \begin{subfigure}[b]{1.0\linewidth}
	\includegraphics[width=1.0\linewidth]{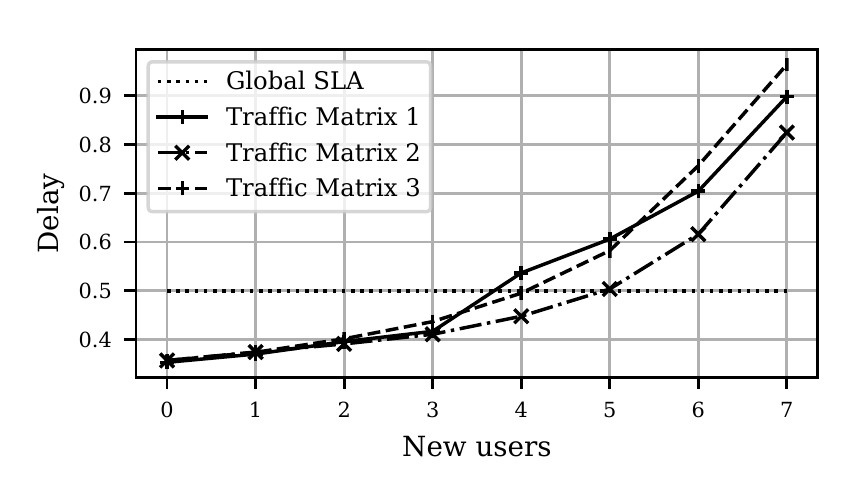}
        \caption{Mean delay}
	\label{fig:mean_users}
    \end{subfigure}
    \begin{subfigure}[b]{1.0\linewidth}
	\includegraphics[width=1.0\linewidth]{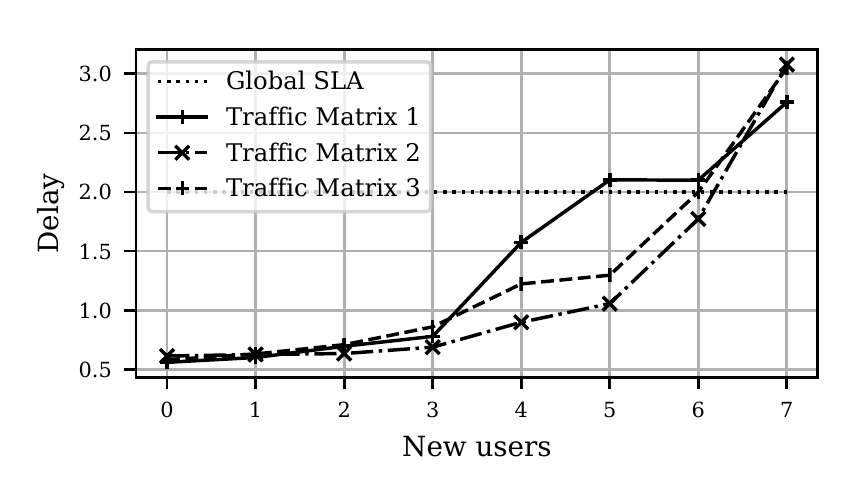}
        \caption{Maximum delay}
	\label{fig:max_users}
    \end{subfigure}
    \caption{Delay optimization as a function of the number of new users}\label{fig:users}
    \vspace{-0.4cm}
\end{figure}

\vspace{0.2cm}
\noindent\textbf{Adding new users}
\vspace{0.1cm}

The objective of this use-case is to evaluate the performance of the network under the presence of potential new users. Each new user in the network increases the amount of traffic that it has to support, and consequently the average and the maximum delay are increased.

Specifically, we explore when certain delay requirements cannot be fulfilled as the number of users with high bandwidth requirements increases. We model these new users as follows: each user multiplies by 2.5 the existing bandwidth demand in a certain node, the first user is connected to node 10, the second one to node 2, the third one to node 8, the fourth one to node 5, the fifth one to node 12, the sixth one to node 1, the seventh one to node 7 and the last one to node 0. We repeat this process under 3 different traffic matrices with initial low traffic intensity (TI=8).

Fig.~\ref{fig:users} shows the mean and maximum delay as new users are subscribed to the network. The dotted line represents the delay requirement, whereas the other lines represent the delay obtained with these different traffic matrices. We observe that the RouteNet model is able to predict the future performance of the network and to know ``a priori'' when the delay requirements will not be accomplished. For example, we observe that a network operating with TM$_1$ will require an update before than the networks operating under the other traffic matrices.

\vspace{0.3cm}
\noindent\textbf{Budget-constrained network upgrade}
\vspace{0.1cm}

In this final use-case, we address a common problem in networking, how to optimally upgrade the network by adding a new link between two nodes. For this, we take advantage of the RouteNet-based model to explore different options to place this new link to select the one that minimizes the mean delay.

Table~\ref{tab:newLink} shows the optimal new placement in the NSF network topology under 10 different traffic matrices with high traffic intensity (TI=15). For each, we also show the average delay before placing the link, the obtained delay with the new optimal link and the delay reduction achieved. We observe that we can achieve an important reduction on the average delay by properly choosing between which nodes this new link is deployed. Note that the optimal placement for the new link depends on the traffic conditions in the network. 

\begin{table}[!h]
\caption{Analysis of the optimal placement of a new link under different traffic matrices: previous delay, delay with the new link and \% of improvement}
\resizebox{\columnwidth}{!}{%
\begin{tabular}{c|c|c|c|c}
\toprule
{\textbf{\begin{tabular}[c]{@{}c@{}}Traffic  \\ matrix\end{tabular}}}&
{\textbf{\begin{tabular}[c]{@{}c@{}}Optimal new  \\ link placement\end{tabular}}} &
{\textbf{\begin{tabular}[c]{@{}c@{}}Previous  \\ delay\end{tabular}}} &
{\textbf{\begin{tabular}[c]{@{}c@{}}Delay with  \\ new link\end{tabular}}} &
{\textbf{\begin{tabular}[c]{@{}c@{}}Delay  \\ reduction\end{tabular}}} \\ 
\midrule
TM$_1$  &  (1, 9)  & 0.732  & 0.478  & 35.7 \%   \\ \hline
TM$_2$  &  (2, 13)  & 0.996  & 0.464  & 53.4 \%   \\ \hline
TM$_3$  &  (1, 9)  & 1.179  & 0.516  & 56.2 \%   \\ \hline
TM$_4$  &  (2, 11)   & 0.966  & 0.518  & 46.37 \%   \\ \hline
TM$_5$  &  (1, 11)   & 0.908  & 0.502  & 44.7 \%   \\ \hline
TM$_6$  &  (0, 13)  & 0.811  & 0.484  & 40.3 \%   \\ \hline
TM$_7$  &  (1, 12)  & 0.842  & 0.485  & 42.4 \%   \\ \hline
TM$_8$  &  (1, 11)   & 0.770  & 0.431  & 44.0 \%   \\ \hline
TM$_9$  &  (1, 9)  & 1.009  & 0.492  & 51.2 \%   \\ \hline
TM$_{10}$  &  (2, 11)   & 1.070  & 0.491 & 54.1 \%   \\ 
\bottomrule
\end{tabular}}
\label{tab:newLink}
\end{table}

\vspace{-0.2cm}
\section{Related work}

Network modeling with deep neural networks is a recent topic proposed in the literature \cite{wangMachineLearning,kdn} with few pioneering attempts. The closest works to our contribution are first Deep-Q \cite{deepQ}, where the authors infer the QoS of a network using the traffic matrix as an input using Deep Generative Models. And second \cite{mestresModeling}, where a fully-connected feed-forward neural network is used to model the mean delay of a set of networks using as input the traffic matrix, the main goal of the authors is to understand how fundamental network characteristics (such as traffic intensity) relate with basic neural network parameters (depth of the neural network). RouteNet is also able to produce accurate estimates of performance metrics -delay and jitter-, but it does not assume a fixed topology and/or routing, rather it is able to produce such estimates with arbitrary topologies and routing schemes not seen during training.  This enables RouteNet to be used for network operation, optimization and what-if scenarios. 

Finally, an early attempt to use Graph Neural Networks for computer networks can be found in \cite{geyer2018learning}. In this case the authors use a GNN to learn shortest-path routing and max-min routing using supervised learning. While this approach is able to generalize to different topologies it cannot generalize to different routing schemes beyond the ones for which has been specifically trained. In addition the focus of the paper is not to estimate the performance of such routing schemes.

\section{Conclusions}

SDN has brought an unprecedented degree of flexibility to network management, which allows the network controller to configure the network behavior up to the flow-level granularity. This flexibility combined with the information provided by network telemetry opens many possibilities for online network optimization.

However, existing network modeling techniques based on analytic models cannot handle this huge complexity. As a result, current optimization approaches are limited to improve a global performance metric, such as network utilization.
Although Deep Learning (DL) is a promising solution to handle such complexity and to exploit the full potential of the SDN paradigm, previous attempts to apply DL to networking problems resulted in tailor-made solutions that failed to generalize to other network scenarios.

In this paper, we presented RouteNet, a new type of Graph Neural Network (GNN) that is specifically designed for modeling computer networks. RouteNet is inspired by the Message-Passing Neural Network (MPNN) previously proposed in the field of quantum chemistry. The main innovation behind RouteNet is a novel message-passing protocol that allows the GNN to capture the complex relationships between the paths and links that form a network topology and the network traffic.

We used RouteNet to model the per-source/destination delay and jitter of a network. Our results show that RouteNet is able to generalize to other network topologies, routing configurations and traffic matrices that were not present in the training set. We finally presented some illustrative use-cases that show the potential of RouteNet to be applied for network optimization in SDN. In particular, we showed that an SDN controller can use RouteNet to optimize multiple KPI and to guarantee the SLA of a particular set of flows, as well as to analyze different what-if scenarios.

\begin{acks}
This work was supported by AGH University of Science and Technology grant, under contract no.~15.11.230.400, the Spanish MINECO under contract TEC2017-90034-C2-1-R (ALLIANCE) and the Catalan Institution for Research and Advanced Studies (ICREA). The research was also supported in part by PL-Grid Infrastructure.
\end{acks}

\bibliographystyle{ACM-Reference-Format}
\bibliography{bibliography}

\end{document}